\documentclass[%
superscriptaddress,
preprintnumbers,
 amsmath,amssymb,
 aps,
prstper,
floatfix,
]{revtex4-2}
\usepackage{graphicx}% Include figure files
\usepackage{dcolumn}% Align table columns on decimal point
\usepackage{bm}% bold math
\usepackage{xcolor}
\begin{document}

\preprint{Preprint}

\title{Role of ambient pressure on bouncing and coalescence of colliding jets}% Force line breaks with \\
% \thanks{A footnote to the article title}%

\author{Minglei Li}
\affiliation{Center for Combustion Energy, Tsinghua University, Beijing 100084, China}%
\affiliation{Department of Mechanical and Aerospace Engineering, Princeton University, Princeton, NJ 08544, USA}%

\author{Abhishek Saha}%
 \email{Email: asaha@eng.ucsd.edu}

\affiliation{Department of Mechanical and Aerospace Engineering, University of California San Diego, La Jolla, CA 92093, USA
}%
\affiliation{Department of Mechanical and Aerospace Engineering, Princeton University, Princeton, NJ 08544, USA}%
\author{Chao Sun}
\affiliation{Center for Combustion Energy, Tsinghua University, Beijing 100084, China}%

\author{Chung K. Law}
\affiliation{Department of Mechanical and Aerospace Engineering, Princeton University, Princeton, NJ 08544, USA}%
\affiliation{Center for Combustion Energy, Tsinghua University, Beijing 100084, China}%

\date{\today}% It is always \today, today,
             %  but any date may be explicitly specified

\begin{abstract}
%\color{blue}
In this letter, the merging-vs-bouncing response of obliquely-oriented colliding jets under elevated and reduced gaseous environment pressures was experimentally examined. Experiments with water and n-tetradecane confirmed that the collision outcome transitions from merging to bouncing, and then to merging again, when the impact velocity was increased. This behavior which was previously reported for atmospheric pressure, has now also been observed at elevated and reduced pressures. New results also show that there exists a critical pressure (0.9 bar for tetradecane and 5 bar for water) below which increasing pressure promotes bouncing (expands the bouncing regime), while beyond this, merging is promoted (reduces the bouncing regime) instead. This leads to a non-monotonic influence of pressure on the non-coalescence outcomes of collisional jets, which was not previously reported. The study provides evidence of new behaviors in colliding jets at reduced and elevated pressures, which differs from well-studied droplet-droplet collisions.
\end{abstract}
\color{black}
%\keywords{Suggested keywords}%Use showkeys class option if keyword
                              %display desired
\maketitle

\begin{figure}[h]
\includegraphics[width=0.75\columnwidth]{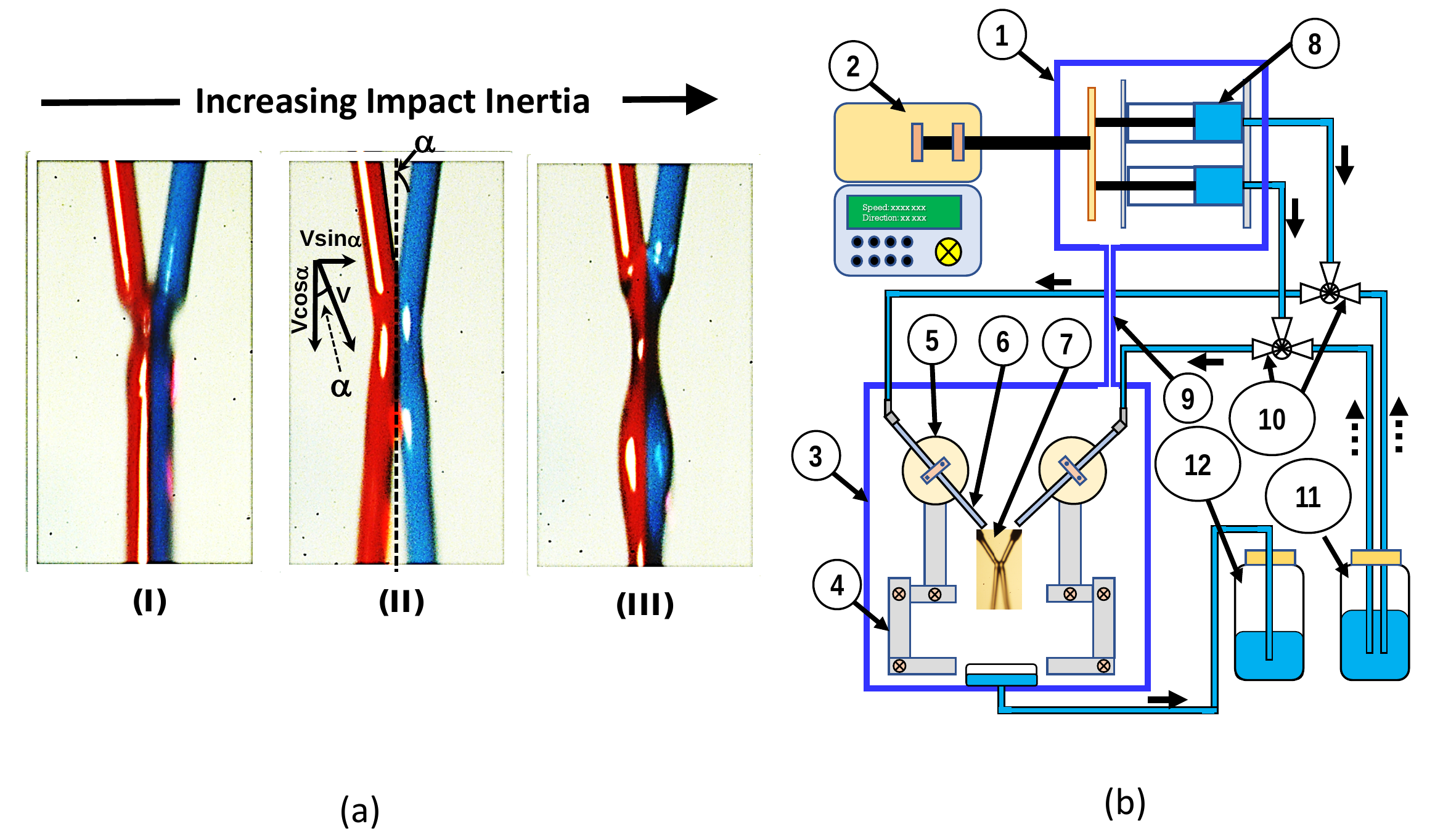}% Here is how to import EPS art
\caption{(a) Images showing the suite of the collision response with increasing collision velocity (or impact inertia). (I): \textit{soft} merging; (II): bouncing; (III): \textit{hard} merging. Liquid: dyed n-tetradecane; pressure: 1 bar.
(b) Schematic illustration of the experimental apparatus. 1. Syringe chamber; 2. Syringe pump controller; 3. Jet chamber; 4. XYZ-micro-position stage; 5. Micro-rotation stage; 6. Jet nozzle; 7. Quartz observation window; 8. Glass syringes; 9. Pressure balance connector; 10. Three-way valves; 11. Liquid reservoir; 12. Waste liquid reservoir.}
\label{fig:snapshots_exp}
\end{figure}

Collision of two liquid masses in gaseous environments is a frequently occurring event in many natural and industrial processes, for example in cloud and raindrop formation as well as spray and mixing processes within liquid-fueled combustors such as diesel and rocket engines \cite{macklin1969subsurface, ching1984droplet,falkovich2002acceleration, roisman2002impact, bouwhuis2012maximal, tran2013air, saha2019kinematics}. 
It has been found that the colliding masses, such as droplet-droplet \cite{jiang1992experimental, qian1997regimes, zhang2011analysis, tang2016dynamics}, droplet-film \cite{pan2007dynamics, tang2016nonmonotonic, Shin-SoftMatter-2020, Wu-POF-2022} and jet-jet \cite{wadhwa2013noncoalescence, li2015dynamics}, could result in non-coalescence and hence bouncing outcomes instead of merging as nominally expected. The cause for the non-coalescence response is the presence of the intervening gas layer between the impacting liquid masses, which can be as thin as tens to hundreds of nanometers \cite{neitzel2002noncoalescence, couder2005bouncing, thoroddsen2012micro, tang2019bouncing} and needs to be ``squeezed'' out by the impacting liquid surfaces before the impact momentum is dissipated. In particular, it has been conclusively demonstrated, for both the droplet and jet systems, that with increasing impact inertia, the collision response can evolve from merging to bouncing to merging again and finally to merging, followed by disintegration of the merged mass. For conceptual visualization, Fig. \ref{fig:snapshots_exp}a shows the experimental images obtained in the course of the present investigation for the suite of possible jet collision outcomes of \textit{soft} merging (I), bouncing (II), and \textit{hard} merging (III), with increasing impact inertia. 
%\color{blue}
Here, \textit{soft} merging denotes merging attained at lower impact velocities, where impact inertia is weak to cause significant deformation of the circular jets, during the impact. \textit{Hard} merging, on the other hand, is attained with high impact velocity or inertia, which causes significant deformation in the shape of the jet \cite{li2015dynamics}. 
\color{black}
Further increase in impact inertia beyond III leads to instabilities that disintegrate the collided jet structure. Such instabilities and the ensuing atomization have been extensively studied in refs. \cite{bush2004collision, bremond2006atomization}.

Recognizing the relevance of system pressure in automotive, airplane, and rocket engines, studies on droplet-droplet collision have also demonstrated that increasing pressure promotes bouncing because of the increased pressure- and hence density-dependent interfacial mass that needs to be displaced. Furthermore, because of the substantial differences in the physical properties between water and hydrocarbons, with water having a larger surface tension and stronger van der Waals force than alkanes, it was also shown that water droplets do not exhibit bouncing at atmospheric pressure, but they readily bounce at higher pressures. On the other hand, while hydrocarbon droplets bounce at atmospheric pressure, they merge at reduced pressures \cite{qian1997regimes}. This property-dependent understanding is enlightening in that extensive investigations on hydrocarbon fuel spray processes for engine applications have inadvertently assumed droplet merging upon collision based on the results of water droplet collision at atmospheric pressure \cite{ORourke-modeling}, and as such have led to biased spray statistics in analyzing engine combustion behavior.  

In view of the above considerations, we have performed a systematic experimental investigation on obliquely oriented colliding jets under both elevated- and reduced-pressure environments. We shall show in due course that the resistance to bouncing for water jets differs from the hydrocarbon jets, as observed for the droplet collision \cite{jiang1992experimental, qian1997regimes}. Furthermore, we have also identified a new phenomenon that, for a certain range of operations, merging instead of bouncing can be promoted at higher pressures, which is contrary to the results of droplet collision \cite{qian1997regimes}.

% The experimental arrangement, results, and analysis are sequentially presented in the following.

% \begin{figure}[h!]
% \includegraphics[width=0.5\columnwidth]{Figures/Exp_fig.pdf}% Here is how to import EPS art
% \caption{Schematic illustration of the experimental apparatus. 1. Syringe chamber; 2. Syringe pump controller; 3. Jet chamber; 4. XYZ-micro-position stage; 5. Micro-rotation stage; 6. Jet nozzle; 7. Quartz observation window; 8. Glass syringes; 9. Pressure balance connector; 10. Three-way valves; 11. Liquid reservoir; 12. Waste liquid reservoir.}
% \label{fig:exp}
% \end{figure}

%\section{Experiments}
To achieve ambient pressure control of the jets, two identical nozzles, together with their relative position and orientation manipulation system, are placed in a sealed chamber with multiple observation windows in specific directions. A schematic of the experimental setup for the controlled-pressure jet collision experiments is shown in Fig. \ref{fig:snapshots_exp}b. %A dual-channel syringe pump keeps the pressure balance between the syringe chamber and the jet chamber, so that the stability of the fluid flux generated from the pump can be guaranteed under different pressure conditions. 
A connected dual-chamber design keeps the pressure balance between the syringe chamber and the jet chamber so that the stability of the fluid flux generated from the pump can be guaranteed under different pressure conditions. 
With this setup, the test liquid is first drawn into the glass syringes from a liquid reservoir and then injected into the nozzles inside the jet chamber to form colliding jets after the chamber pressure is established. The flow rate of the system can be varied from 1.0 to 28.0 ml/min with an accuracy of ±0.5\%. Considering the extra pressure resistance from the pipes, adapters, and nozzles, the flux is calibrated under various pressure conditions, according to which the experimental results are further amended to keep the data accurate. Two identical nozzles with an inner diameter of 200 µm were designed by fixing the short thin-wall capillary tube inside a drilled copper column. Compared with the normally used glass capillary or needles, such nozzles yield sharper edges to avoid tip adhesive attraction of the liquid without increasing the pressure resistance and hence are able to generate fluid jets with a larger velocity range. The impact angle ($\alpha$, shown in Fig. \ref{fig:snapshots_exp}a) and position of the nozzles are precisely manipulated through micro-rotation and XYZ-micro-positioning, which can be conveniently controlled outside the chamber through soft driver connectors. Technical grade de-ionized water and n-tetradecane are used as the working fluids; their properties are listed in Table~\ref{tab:property}. %The bouncing-merging transition experiments at 1 bar (atmospheric) condition were performed on all the alkanes, while for experiments at elevated and reduced pressure, water and n-tetradecane were selected.

 \begin{table}[h]%The best place to locate the table environment is directly after its first reference in text
\caption{Properties of the liquids used for the study}
\label{tab:property}
\begin{ruledtabular}
\begin{tabular}{cccc}
\textrm{Liquid}&
\textrm{density, $\rho$ (kg/m$^3$)}&
\multicolumn{1}{c}{\textrm{dynamic viscosity, $\eta$ ($10^{-3}$Pa-s)}}&
\textrm{surface tension, $\sigma$ ($10^{-3}$N/m)}\\
\colrule
Water (H$_2$O) & 1000 & 1.01 & 72.90\\
%n-Decane & 730 & 0.92 & 23.83\\
%n-Dodecane & 750 & 1.34 & 25.35\\
n-Tetradecane (C14) & 760 & 2.30 & 26.56\\
%n-Hexadecane & 773 & 3.00 & 27.47\\
\end{tabular}
\end{ruledtabular}
\end{table}

\begin{figure}[h!]
\includegraphics[width=0.70\columnwidth]{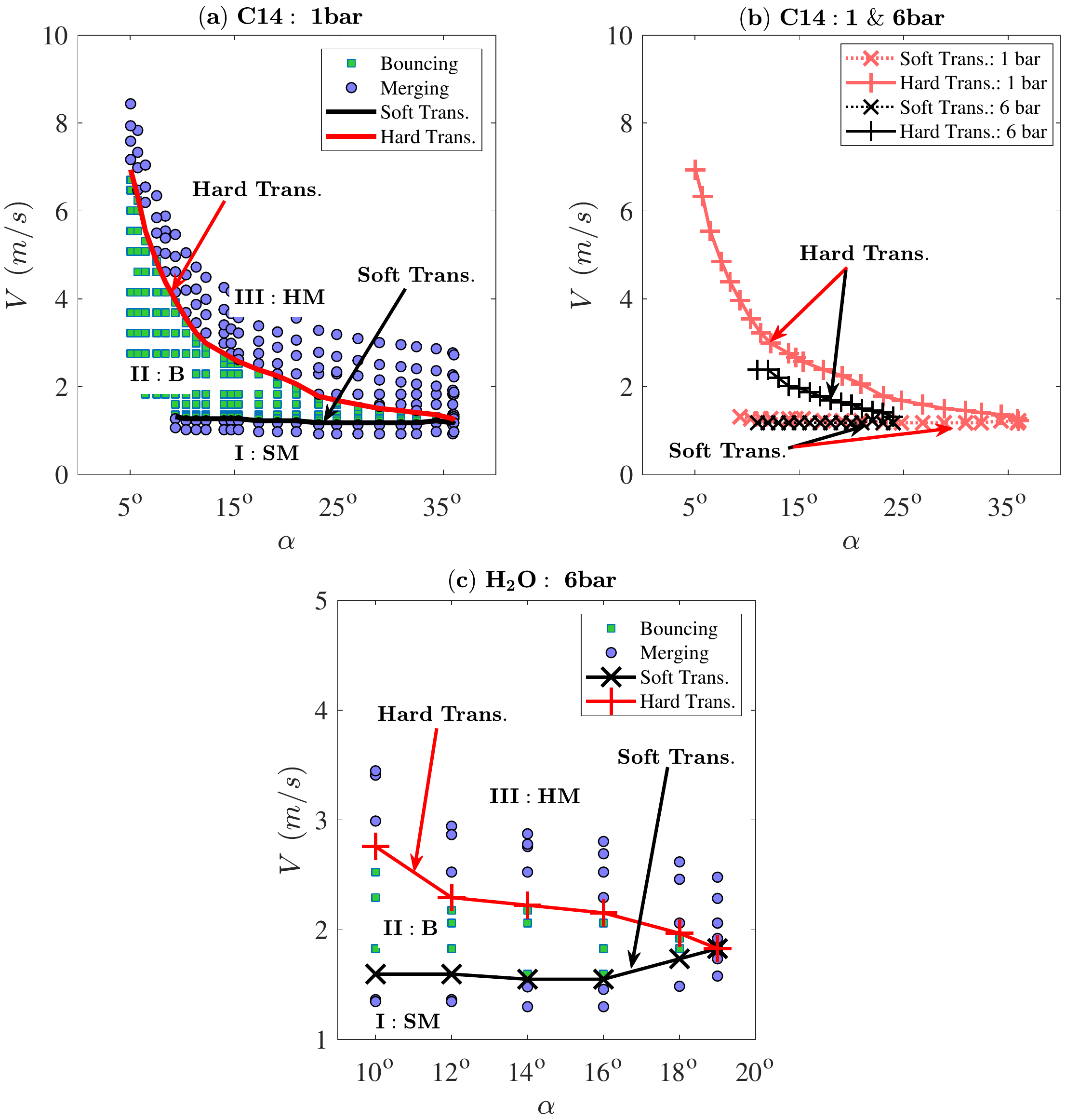}% 
\caption{(a) Regime map for n-tetradecane (C14) at 1 bar. (b) Comparison of regime boundaries for n-tetradecane (C14) at 1 bar and 6 bar. (c) Regime map for water (H$_2$O) at 6 bar. Various impact outcomes: bouncing, (\textit{soft} and \textit{hard}) merging and the (\textit{soft} and \textit{hard}) transition boundaries are shown. The data for n-tetradecane at 1 bar was presented in \cite{li2015dynamics}. The uncertainty in $V$ is about $\pm5\%$. \textbf{SM}: soft merging; \textbf{B}: bouncing; \textbf{HM}: hard merging.}
\label{fig:angle_map}
\end{figure}

% \section{Results and Discussion}

The bouncing behavior of n-tetradecane under the normal pressure (1 bar) condition is first investigated. By gradually varying the jet velocity and impact angle, we have mapped the regimes within which the collision yields (soft) merging, bouncing, and (hard) merging, with the bouncing outcome separating the soft and \textit{hard} merging outcomes. Figure \ref{fig:angle_map}a maps the merging-vs-bouncing response of the colliding jets and shows the dependence of critical velocity delineating the transition with the collisional angle of the jets. The critical velocities generate two boundaries for \textit{hard} and \textit{soft transitions}, splitting the outcome into regimes I, II, and III. 
%\hl{Define soft and \textit{hard} merging}
It is seen that the \textit{soft} merging transition boundary is nearly insensitive to impact angle ($\alpha$), having an almost constant critical velocity, while the \textit{hard} merging transition boundary has a strong dependence on the impact angle, with a significant decrease in the critical velocity. 
These two transition boundaries merge at a certain angle ($\alpha_{cr}$), 38\textdegree~for n-tetradecane, beyond which bouncing is not observed. Similar transition boundaries are obtained for other alkanes, n-hexadecane, n-dodecane and n-decane, with $\alpha_{cr}$ being 44\textdegree, 28\textdegree ~and 19\textdegree ~respectively, as reported in our previous work \cite{li2015dynamics}. 
In that work, we analyzed and showed that the \textit{hard transition} boundary is dominated by the gas-layer thickness ($H_d$) evolution which scales as 
$H_d/R=A_I St^{-2/3}$, where $St=(\rho_l RU)/ \eta_g$ is the Stokes number, $A_I$ a prefactor, $\rho_l$ the liquid density, $R$ the jet radius, $\eta_g$ the air viscosity and $U$ the effective impact velocity which is $Vsin\alpha$ in the present case. 
The \textit{soft transition} boundary is dominated by the Plateau-Rayleigh instability of the jet surface, leading to the velocity ratio, $\Gamma=V/V_c$, where $V_c=\sqrt{\sigma/(\rho_lR)}$ is the velocity of the capillary waves traveling along the jet and $\sigma$ is surface tension.

Since the gas density can directly influence the inertia of the gas layer, it is straightforward to consider its effects by varying the gas pressure. First, to assess whether the pressure variation can affect the liquid side of the bouncing jet, we captured the jet shape and measured the characteristic geometry parameters, including the contact length, maximum deformation, and separation angle of the bouncing jets, respectively, under pressures from 1 to 6 bar. Results show that the shape evolution of the liquid jets has no visible dependence on the system pressure, which implies that, in this case, the ambient pressure variation has no effect on the liquid side during the jet bouncing process. This is to be expected because the viscosity and surface tension of the liquid barely change within this pressure range, and the density ratio between liquid and air, $\rho_l/\rho_g$, is over 500, which is much larger than the gas density variation. Consequently, the influence of pressure on the non-coalescence behavior of the jets could only play a role through the gas layer dynamics.

%According to the widely investigated non-coalescence phenomenon of head-on droplet collision [19–21], while no bouncing is observed at normal atmosphere pressure for water droplets, they exhibit bouncing at higher pressures. On the contrary, while colliding n-tetradecane droplets would bounce back at the normal pressure, they would merge if the pressure is low enough. Furthermore, the bouncing regime of n-tetradecane droplets is substantially enlarged in pressurized circumstance and shrink in reduced pressures. This is because the intervening gas layer becomes denser and thereby harder to be squeezed out with increasing ambient pressure, making droplets easier to bounce back, and vice versa if the pressure decreases. [Delete this paragraph as it seems to repeat what has been stated before.]

\begin{figure}[t]
\includegraphics[width=0.70\columnwidth]{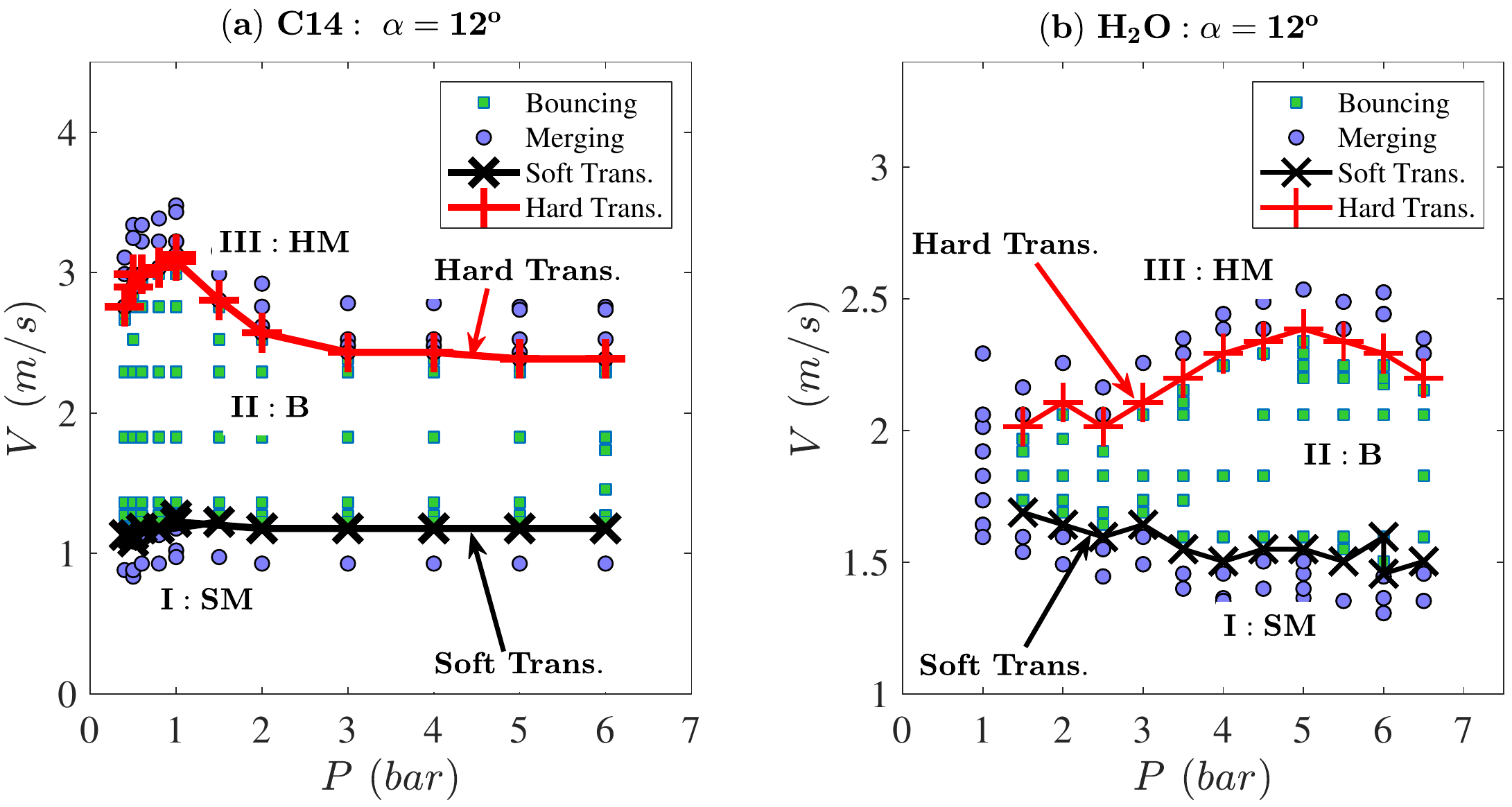}% 
\caption{Regime map as functions of the impact velocity and ambient pressure for (a) n-tetradecane (C14) and (b) Water (H$_2$O). Impact angle, 
$\alpha$=12\textdegree. The uncertainty in $V$ is about $\pm5\%$. \textbf{SM}: soft merging; \textbf{B}: bouncing; \textbf{HM}: hard merging.}
\label{fig:Press_map}
\end{figure}

Subsequent extensive experimental investigations on jet collision in elevated and reduced pressure ambiance were conducted, showing that, similar to droplet collision, no bouncing is observed for water jets under normal pressure (1 bar), and is exhibited at higher pressures, such as 6 bar as shown in Fig. \ref{fig:angle_map}c. For n-tetradecane jets, bouncing is observed at normal pressure but is suppressed when the pressure is reduced to 0.2 bar. 
The investigation, however, identified an important qualitative difference between droplet and jet collisions at high pressures. With increasing impact inertia and within the range of higher pressures investigated, the \textit{hard transition} boundary, from bouncing to \textit{hard} merging, is reduced instead of expanded for n-tetradecane. 
%\color{blue}
As shown in Fig. \ref{fig:angle_map}b, the \textit{hard transition} velocities at 6 bar are much smaller than those at 1 bar, leading to a narrower bouncing regime and rendering the critical collisional angle decreasing from 38\textdegree ~to 24\textdegree.
\color{black}
To further explore and quantify this new phenomenon, additional experiments were conducted with water and n-tetradecane for various pressures while keeping the jet collision angle fixed at 12\textdegree. Figure \ref{fig:Press_map} quantifies the non-monotonic behavior of the \textit{hard transition} boundary with increasing pressure, first increasing and then decreasing, and hence promoting and inhibiting bouncing, respectively. 
%\color{blue}
For water jets, a small change in the \textit{soft transition} boundary is also observed, in that it decreases slightly with increasing pressure (Fig. \ref{fig:Press_map}b). However, the non-monotonicity in \textit{hard transition} boundary is more dramatic, and it controls the expansion/reduction in the bouncing regime.
\color{black}
The transition state (from expansion to reduction in bouncing regime) occurs at a critical pressure $P_{cr}$ of 0.9 bar for n-tetradecane (Fig. \ref{fig:Press_map}a) and 5 bar for water (Fig. \ref{fig:Press_map}b).
%\color{blue}
Limited experiments with other impact angles at various pressures were also conducted. Although they were insufficient to draw complete regime maps at other impact angles, we observed similar behavior reported in Fig. \ref{fig:Press_map}.
\color{black}

% To identify the cause of this non-monotonic response that the bouncing jets can keep stable is that the intervening gas layer between the liquid interfaces is continuously being supplemented by upstream air entrainment, which is the major difference from droplet collision. Since the increasing pressure cannot influence the liquid deformation, as mentioned earlier, further analysis shows that increasing ambient pressure can also reduce the extent of the gas entrainment by reducing the air boundary layer thickness. Specifically, in analogy with the boundary layer flow over a plate, the air boundary layer thickness over the jet surface would scale as $\delta \sim l/Re^{1/2}$, where $Re=\rho_g V l/\eta_g$ is the je Reynolds number, $l$ the free jet length, and $\rho_g$ and $\eta_g$ the gas density and viscosity coefficient respectively. Since $\rho_g$ increases with pressure, while $\eta_g$ is insensitive to it, $\delta$ will decrease with increasing pressure. This reduced boundary layer thickness not only counteracts the increased mass entrainment due to the increased pressure, but it also reduces the interfacial separation distance and hence promotes the attractive van der Waals force. Consequently, the observed non-monotonic trend is due to the joint influence of pressure variation on gas layer properties, gas density and entrainment. The coalescence-vs-bouncing responses of droplet-droplet and jet-jet collision can therefore be interpreted from a unified viewpoint.

In summary, we have identified the pressure effect on the jet-jet collision response. Unlike the widely investigated droplet-droplet collision case, pressurized ambient gas does not always promote bouncing in jet-jet collision. A unified non-monotonic trend of bouncing transition boundary is identified, leading to a critical pressure under which the collisional jets achieve maximum bouncing. The understanding gained, herein, yields a useful reference for practical applications, offering an additional consideration to the role of pressure on the bouncing versus coalescence of colliding jets. The experimental observations also serve as benchmark data for future theoretical and modeling expositions, which are required to provide additional details on the underlying interfacial dynamics.

% \section*{Authors' Declarations}
% The authors have no conflicts to disclose.
% \section*{Data Availability Statement}
% The figures in the manuscript contain most of the data collected for this study.

\bibliography{apssamp}% Produces the bibliography via BibTeX.

\end{document}